\documentclass[twocolumn,aps]{revtex4}
\input{epsfig.sty}
\newcommand{\beq}{\begin{eqnarray}}
\newcommand{\eeq}{\end{eqnarray}}
\begin{document}
\title{Electromagnetic field creation during EWPT nucleation with MSSM}
\author{Ernest M. Henley\\
Department of Physics, University of Washington, Seattle, WA 98195\\
Mikkel B. Johnson \\
Los Alamos National Laboratory, Los Alamos, NM 87545 \\
Leonard S. Kisslinger\\
Department of Physics, Carnegie-Mellon University, Pittsburgh, PA 15213\\ }

\begin{abstract}
   We derive the equations of motion for electroweak MSSM with a right-handed
Stop, from which we derive the equations for the electromagnetic field that 
arises from bubble nucleation and collisions during the first order electroweak
phase transition that can occur in this MSSM. Introducing an isospin ansatz
we derive e.o.m. for the electrically charged W fields uncoupled from all 
other fields. These serve as the current for the Maxwell-like e.o.m. for the
em field. The resulting electromagnetic field arising during EWPT bubble 
nucleation is found.  This electromagnetic structure, along with that arising 
from bubble collisions, could seed galactic and extra-galactic magnetic fields.

\end{abstract}
\maketitle
\noindent
PACS Indices:12.38.Lg,12.38.Mh,98.80.Cq,98.80Hw
\vspace{1mm}


\vspace{1mm}
\noindent
hep-ph/0503247

\section{Introduction}

The origin of the large-scale galactic and extra-galactic 
magnetic fields which have been observed (see Ref\cite{gr} for a review)
is a long-standing problem of astrophysics. There has been a great interest
in possible cosmological seeding of these magnetic fields in the early 
universe phase transitions: the
Quantum Chromodynamics chiral phase transition (QCDPT) from the quark-gluon
plasma to our hadronic universe, and the electroweak phase transition
(EWPT) in which the Higgs and the other particles acquired their masses.
Research on the QCDPT has explored not only seeding of the galactic and
extra-galactic structure, but also possible effects in Cosmic Microwave
Background Radiation (CMBR) correlations\cite{co,soj,kl,bfs,dfk,fz,kv95,ae98,
cst00,son99}. In our own work we have explored
possible large scale magnetic structure which could produce observable 
CMBR polarization correlations\cite{lsk1} arising from bubble collisions
during the QCDPT\cite{lsk2}.

The EWPT is particularly interesting for exploring possible cosmological
magnetic seeds since the electromagnetic (em) field along with the 
$W^{\pm}$ and $Z$ fields are the gauge 
fields of the Standard model, while in the QCDPT the em field is included
in the Lagrangian through coupling to quarks. It is now believed that 
in the Standard model there is no first order EWPT\cite{klrs}, no
explanation of baryogenesis, or any interesting cosmological magnetic
structures created during the crossover transition. However, there has
been a great deal of activity in the supersymmetric extension of the 
standard model\cite{r90}, and with a MSSM having a Stop with a mass 
similar to the Higgs there can be a first order phase transition and 
consistency with baryogenesis\cite{laine,cline,losada}. 

In the present work we use the MSSM with the form of the Standard EW 
Model plus a right-handed Stop, which has been used by a number of 
authors to examine the EWPT and
baryogenesis. See, e.g., Refs. \cite{laine,bodeker}. There have been other
models for CP violation and baryogenesis, such as two-Higgs models (see
Refs.\cite{laine,cline,losada} for references and discussion) and
leptoquarks (see, e.g., Ref.\cite{herczeg} for a discussion and references).
For the present exploratory work on nucleation we could use, e.g., a two-Higgs
model, as the precise nature of the extension of the Standard EW Model
is not needed. For future work on magnetic field generation with possible
bubble collision we shall use the MSSM and treat all the equations of
motion, derived below, with a right-handed Stop.

In the most detailed previous research on the study of magnetic fields 
arising from EWPT transitions\cite{kv95,ae98,cst00}  a model was used in 
which the equations of motion (e.o.m.) for the em field involved the chargeless
Higgs. The EWPT bubbles were empty of the em field until the bubbles 
collided and overlapped. In the present work we derive the e.o.m.
directly from the EW Lagrangian, using a MSSM including the right-handed Stop 
field for consistency with a first order phase transition. The e.o.m. which 
we obtain for the em field is Maxwell-like, with the current given by the
charged $W^{\pm}$ fields, which is physically reasonable. Fermions are not
considered in the present work. 

In general, 
the e.o.m. are complicated coupled partial differential equations,
in which solutions for the Higgs and Stop fields must be found to
obtain the current for the em field equation. In the present paper, however, 
we introduce an I-spin formulation which allows us
to uncouple the e.o.m. for the $W^{\pm}$ fields, and carry out a study of
spherically symmetric EW bubble nucleation. For pure nucleation of a
bubble we solve the $W^{\pm}$ e.o.m. numerically, and obtain solutions for
the em field using a fit to these numerical solutions for the $W^{\pm}$
fields. The solutions have an instanton-like form near the bubble wall,
as expected.

  These are the first solutions for the electromagnetic field
from EWPT nucleation that have been obtained from the EW Lagrangian.
Due to spherical symmetry during nucleation only electric fields are
produced.  These give the initial fields for bubble collisions.
For the derivation of EW cosmological magnetic seeds, the
more complicated collision problem must also be solved, which we shall
attack in the near future, and fermion fields will be included.

   In Sec. II we review the previous work on electromagnetic field 
creation during the EWPT.
In Sec. III we derive the equations of motion with the MSSM,
and in Sec. IV give the form with our I-spin ansatz. In Sec. V
we discuss our method for finding solutions, and give our results.
In Sec. VI we give a summary of the paper and conslusions.

\section{Previous treatment of electromagnetic field
creation during a first-order EWPT} 

   The most detailed derivation of the creation of magnetic fields during the
EWPT is based on the Abelian Higgs model introduced by Kibble and
Vilenkin\cite{kv95}, with the Lagrangian
\beq
\label{kv}
      {\cal L}^{AH} &=& -\frac{1}{4}F_{\mu\nu}F^{\mu\nu} +(D_\mu \Phi)^\dagger
 D^\mu \Phi +V(\Phi) \; ,
\eeq
where $D_\mu = \partial_\mu +ieA_\mu$, with $A_\mu, F^{\mu\nu}$ the electromagnetic
4-potential, tensor, and $\Phi$ is the complex Higgs field,
\beq
\label{Higgs}
              \Phi &=& \frac{1}{\sqrt{2}} \rho e^{i\theta} \; .
\eeq

The Higgs potential, $V(\Phi)$, has two minima, and as in Ref\cite{col77} it
is assumed that there is a first order phase transition corresponding to
the transition from the false to the true vacuum. In the model used in
Refs\cite{kv95,ae98,cst00} $\rho$ is assumed to be a constant corresponding
to the Higgs mass. In this model, with the only electric charges that are
present being associated with the gauge derivative $D_\mu$,  the
electromagnetic current is given by
\beq
\label{emcurrent}
                  j_\mu &=& i e \Phi^\dagger D_\mu \Phi + c.c. \nonumber \\
                        &=& -e \rho^2 (\partial_\mu \Phi +A_\mu) \; .
\eeq
Futhermore it is assumed that there is uniformity during the nucleation of
the EW bubbles during the first-order phase transition, leading to nucleating
bubbles empty of electromagnetic fields. Magnetic fields are created only
during collisions of bubbles within the region of overlap of the bubbles.

  As discussed in the Introduction,  one problem with the standard model for 
producing magnetic seeds is that there is no first order phase transition and
therefore no bubbles.   With the introduction of a right-handrd Stop, giving
the MSSM that we use in the present paper, there can be a first order phase
transition, and bubbles are produced, as assumed in the Kibble-Vilenkin model.
Moreover, with this MSSM based on the standard EW model, the surfaces of
the bubbles are composed, in part, from the charged gauge fields, and therefore
create electromagnetic fields from bubble nucleation before collisions. 
We therefore find electromagnetic fields from the nucleation, before
collisions. This will produce new boundary conditions for the creation of
magnetic fields during the bubble collisions.

   It is interesting to note that in the numerical calculations of
 Ref\cite{cst00} the Higgs field grows at the surface of the nucleating
bubble wall to resemble the instanton-like solutions that we find.
Also, we should add that in our work on EWPT collisions producing magnetic
fields (in progress) we have found some of the approach of Ref\cite{ae98}
very useful.

\section{MSSM EW equations of motion with right-handed 
stop}

In this section we derive the equations of motion for the
standard Weinberg-Salam model, in the electroweak MSSM with all
partners of the standard model fields integrated out except the
Stop, the partner to the top quark.

\beq
\label{L}
  {\cal L}^{MSSM} & = & {\cal L}^{1} + {\cal L}^{2}  + {\cal L}^{3} 
\nonumber  \\
      && +{\rm leptonic \: and \: quark \: interactions }\\ \nonumber
         {\cal L}^{1} & = & -\frac{1}{4}W^i_{\mu\nu}W^{i\mu\nu}
  -\frac{1}{4} B_{\mu\nu}B^{\mu\nu} \\ \nonumber
 {\cal L}^{2} & = & |(i\partial_{\mu} -\frac{g}{2} \tau \cdot W_\mu
 - \frac{g'}{2}B_\mu)\Phi|^2  -V(\Phi) \nonumber \\
 {\cal L}^{3} &=& |(i\partial_{\mu} -\frac{g_s}{2} \lambda^a C^a_\mu)\Phi_s|^2
    -V_{hs}(\Phi_s,\Phi) \nonumber \, ,
\eeq
where the pure $C^a_\mu$ term is omitted in ${\cal L}^{1}$ and  
\beq
\label{wmunu}
  W^i_{\mu\nu} & = & \partial_\mu W^i_\nu - \partial_\nu W^i_\mu
 - g \epsilon_{ijk} W^j_\mu W^k_\nu\\ \nonumber
 B_{\mu\nu} & = & \partial_\mu B_\nu -  \partial_\nu B_\mu \, ,
\eeq
where the $W^i$, with i = (1,2), are the $W^+,W^-$ fields, $C^a_\mu$
is an SU(3) gauge field, ($\Phi$, $\Phi_s$) are the (Higgs, right-handed 
Stop fields), $(\tau^i,\lambda^a)$ are the (SU(2),SU(3) generators, and
the electromagnetic and Z fields are defined as 
\beq
\label{AZ}
   A^{em}_\mu &=& \frac{1}{\sqrt{g^2 +g^{'2}}}(g'W^3_\mu +g B_\mu) \nonumber \\
   Z_\mu &=& \frac{1}{\sqrt{g^2 +g^{'2}}}(g W^3_\mu -g' B_\mu) \; .
\eeq 
The effective Higgs and Stop potentials are taken as
\beq
\label{V}
    V(\Phi) & = & -\mu^2 |\Phi|^2 + \lambda |\Phi|^4 \nonumber \\
    V_{hs}(\Phi,\Phi_s) & = & -\mu^2_s |\Phi_s|^2 + \lambda_s |\Phi_s|^4 \\
              &&  +\lambda_{hs} |\Phi|^2 |\Phi_s|^2 \, .\nonumber
\eeq
The various parameters are discussed in many publications\cite{laine}. In 
particular we
need $g=e/sin\theta_W = 0.646$, $g'=g \; tan\theta_W =0.343$, and 
$G=gg'/\sqrt{g^2 +g^{'2}}=0.303$.

In the picture we are using, the Higgs and Stop fields will play a dynamic
role in the EW bubble nucleation and collisions, and we shall 
need the space-time structure of these fields rather than only
the vacuum expectation value for a particular vacuum state for
the complete solutions of the e.o.m.
Our form for the Higgs field, $\Phi$, is 
\beq
\label{phi}
         \Phi(x) & = & \left( \begin{array}{clcr} 0 \\
                                \phi(x)
         \end{array} \right) \; .
\eeq
and 
\beq
\label{tau}
  \tau \cdot W_\mu \Phi & = & 
    \left( \begin{array}{clcr} (W^1_\mu-iW^2_\mu) \\
                            - W^3_\mu
         \end{array} \right)  \phi(x) \, .
\eeq

In the present exploratory paper treating bubble nucleation
we center on the possible generation of an electromagnetic field,
and the solution of all of the e.o.m. is avoided. 
Therefore, specific forms and solutions for the Higgs and Stop fields
do not enter the equations needed for the present
work. For this reason we do not choose a specific form for the right-handed
Stop field, $\Phi_s$, and for convenience write the additional MSSM gauge
field as
\beq
\label{cmu}
              C_\mu &=& \frac{\lambda^a}{2} C^a_\mu \; .
\eeq
We also use the definitions
\beq
\label{phis}
       \phi(x) &\equiv& \rho(x)e^{i\Theta(x)} \nonumber \\
       |\phi(x)|^2 &=& \rho(x)^2 \\ 
        |\Phi_s(x)|^2 & \equiv & \rho_s(x)^2 \nonumber \; .
\eeq         
Although we do not need specific forms for  $C_\mu$ or $\Phi_s$,
we assume that a Stop condensate is formed for consistency with a first order
EWPT, as in Ref \cite{bodeker}

With these definitions $ {\cal L}^{2}$ is (j = (1,2,3)) 
\beq
\label{L2}
 {\cal L}^2 & = &  \partial_\mu \phi^*\partial^\mu \phi +
[i(\partial_\mu \phi^*)\phi - i\phi^* \partial_\mu \phi]
( -g W^3_\mu \nonumber \\
          && +g' B_\mu)/2 + \phi^* \phi[(\frac{g}{2})^2 (W^j)^2 +
(\frac{g'}{2})^2 B^2 \nonumber \\
 &&-\frac{gg'}{2} W^3 \cdot B]-V(\phi) \; .
\eeq 

The equations of motion are obtained by minimizing the action
\beq
\label{action}
 \delta \int d^4 x [{\cal L}^{1}+{\cal L}^{2} +{\cal L}^{3}] & = & 0 \, ,
\eeq
i.e., we do not include the leptonic parts of ${\cal L}$.
The equations of motion that we obtain from the variations in 
$W^{1}$, for i=(1,2) are
\beq
\label{eom1}
  && \partial^2 W^i_\nu-\partial^\mu \partial_\nu  W^i_\mu
 -g \epsilon^{ijk} {\cal W}^{jk}_\nu
 +\frac{g^2}{2}\rho(x)^2 W^i_\nu \nonumber \\
     && \;\;\; = 0 \; ,
\eeq
with 
\beq
\label{w's}
     {\cal W}^{jk}_\nu &\equiv&  \partial^\mu( W^j_\mu) W^k_\nu  
     + W^j_\mu \partial^\mu  W^k_\nu + W^{j\mu} W^k_{\mu\nu}\;, \nonumber \\
     W^k_{\mu\nu}&=& \partial^\mu W^k_\nu-\partial^\nu W^k_\mu-g\epsilon^{klm}
  W^l_\mu W^m_\nu
\eeq
The e.o.m. for $A^{em},Z$ are
\beq
\label{eom2}
 &&\partial^2 A^{em}_\nu-\partial_\mu \partial_\nu  A^{em}_\mu
 -\frac{gg'}{\sqrt{g^2 + g^{'2}}} \epsilon^{3jk} {\cal W}^{jk}_\nu \nonumber \\
 &&\;\;\;\;\; = 0
\eeq
\beq
\label{eom3}
 &&\partial^2 Z_\nu-\partial_\mu \partial_\nu Z_\mu -
\frac{\rho^2\partial_\nu \Theta}{\sqrt{g^2 + g^{'2}}} \nonumber \\
&&  -\frac{g^2}{\sqrt{g^2 + g^{'2}}} \epsilon^{3jk} {\cal W}^{jk}_\nu 
  = 0\;.
\eeq

 The e.o.m. for the Higgs field are
\beq
\label{eom4}
 &&\frac{1}{\rho(x)} \partial^2 \rho(x) -\mu^2+ 2\lambda \rho(x)^2 +
\lambda_{hs} \rho_s(x)^2  -H \cdot H \nonumber \\
&&-\partial_\mu \Theta \partial^\mu \Theta 
+\frac{\sqrt{g^2 + g^{'2}}}{2} Z^\mu \partial_\mu \Theta = 0 \; ,
\eeq
with
\beq
   H \cdot H &\equiv&  
  (\frac{g}{2})^2 W^i \cdot W^i
+(\frac{g'}{2})^2 B \cdot B -\frac{gg'}{2} W^3 \cdot B \nonumber\; ,
\eeq
and
\beq
\label{eom5}
 \partial_\mu (\rho(x)^2 \partial^\mu \Theta -\frac{\sqrt{g^2 + g^{'2}}}{2}
  \rho(x)^2 Z^\mu ) &=& 0 \; .
\eeq
The e.o.m. for the right-handed Stop is 
\beq
\label{stop}
  &&-\partial^2 \Phi_s +ig_S(\partial^\mu(C_\mu\Phi_s)+(C_\mu\partial^\mu\Phi_s))
 +(g_s^2 C_\mu^\dag C_\mu \nonumber \\
  && +\mu_s^2+2\lambda_s\rho_s^2 +\lambda_{hs}\rho^2)
\Phi_s = 0 \; ,
\eeq
We do not give the e.o.m. for the $C_\mu$ gauge field, which is not needed in 
the present work.

These are exact equations of motion in our MSSM model with a right-handed stop.
Note that in the absence of external currents there are trivial solutions
to the e.o.m. for the gauge fields, with all A=Z=W=0, just as in pure
electrodynamics, but the solutions with nonvanishing gauge fields are
the ones of physical interest. Moreover, since the EW fields are present
in the universe before EWPT bubble nucleation, from the boundary
conditions, they must be nonvanishing as bubble nucleation begins.
From our e.o.m. we derive the gauge fields during nucleation of the
bubbles, including the electromagnetic field.

\section{I-Spin Ansatz and Electromagnetic Field Creation from $W^{\pm}$}

   One of the most important features of the equations of motion derived
directly from the EW Lagrangian is that the source current  of the 
electromagnetic field is given by the charged gauge $W^{\pm}$ fields,
as seen from the Maxwell-like Eq.(\ref{eom2}). Although this differs from
the Kibble-Vilenkin\cite{kv95}, Ahonen-Enqvist\cite{ae98}, 
Copeland-Saffin-T$\ddot{o}$rnkvist\cite{cst00} picture, much of the underlying 
physics is the same. Note that our equations are charge symmetric, so there 
is no net electric charge produced and charge is conserved, but as explained 
above, individual bubbles have $W^{\pm}$ constituents, which produce 
electromagnetic fields during nucleation.

This suggests that we use an SU(2), isospin ansatz for the gauge 
fields. We assume in the present paper that 
\beq
\label{nuci1}
  W^j_\nu &\simeq& i\tau^j W_\nu(x) \simeq i\tau^j x_\nu W(x)\;\; {\rm j=1,2,3}
 \nonumber \\
  A^{em}_\nu &\simeq& i\tau^3 A_\nu(x)\simeq i\tau^3 x_\nu A(x) \nonumber \\
  Z_\nu &\simeq& i\tau^3 Z(x)_\nu \simeq i\tau^3 x_\nu Z(x) \; , 
\eeq
with the I-spin operators defined as $\epsilon^{mjk} \tau^j\tau^k = i\tau^m$.
We shall see that this enables us to derive the straight-forward
equations of motion for the electromagnetic field, which can be solved
to a good approximation for symmetric nucleation of EW bubbles. In this 
section we derive the e.o.m. for spherically symmetric bubble nucleation, 
so that $W(x)=W(r,t)$ and $A(x)=A(r,t)$,
with $x^\mu x_\mu = t^2-r^2$. First note that
\beq
\label{Wjk}
     \epsilon^{ijk} {\cal W}^{jk}_\nu &=& i\tau^i \times F[W_\nu,
\partial_\nu W]\; ,
\eeq
with $F$ a function of $W_\nu$ and $\partial_\nu W$ to be 
determined, so that the e.o.m. for $W_\nu$, Eq.(\ref{eom1}), becomes
\beq
\label{eom6}
  && \partial^\mu \partial_\mu W_\nu-\partial_\nu \partial_\mu W^\mu
-gx_\nu [5 W^2+  \nonumber \\
 && 3 W(t\partial_t+ r\partial_r)W +g s^2 W^3  -\beta s^2 \partial_r W]
\nonumber \\ 
 && -\frac{g^2}{2} \rho^2 W_\nu = 0 \; ,
\eeq
with $W_\nu = x_\nu W(r,t),\; s^2 = t^2-r^2,\; r=\sqrt{\sum_{j=1}^{3} x^j x^j}
= \sqrt{-\sum_{j=1}^{3} x^j x_j},\; \partial_j r= x^j/r$, and $\beta  = (+,-)$ 
for $\nu = (t,j)$. Subtracting 
the e.o.m. for $W_t \times x_j$ from the e.o.m. for $W_j \times t$ we find 
\beq
\label{eomW}
  && (\partial_t^2 + \partial_r^2)W +\frac{t^2+r^2}{rt}\partial_t
 \partial_r W +(\frac{3}{r}\partial_r + \frac{3}{t}\partial_t)W \nonumber \\
  &&+gW (t^2-r^2)(\frac{1}{t}\partial_t-\frac{1}{r}\partial_r)W = 0 \\
\label{eomA}
   && (\partial_t^2 + \partial_r^2)A +\frac{t^2+r^2}{rt}\partial_t
 \partial_r A +(\frac{3}{r}\partial_r + \frac{3}{t}\partial_t)A \nonumber \\
  && +GW (t^2-r^2)( \frac{1}{t}\partial_t-\frac{1}{r}\partial_r)W = 0 \; .
\eeq

The most significant aspect of the I-spin formulation is that we obtain
e.o.m. for $W(r,t)$ and for $A(r,t)$ without contributions from the Higgs 
or Stop fields, because they decouple. Although the Stop and Higgs fields
disappear from these equations, however, they are essential to
obtain a first-order EWPT and EW bubbles, as discussed in the Introduction.

As pointed out at the beginning of this section, the current for the 
electromagnetic field arises entirely form the electrically charged 
fields/particles, $W^{\pm}$, as seen also in Eq.(\ref{eomA}). Moreover,
the current within our I-spin formulation is determined by a nonlinear
partial equation for $W(r,t)$, without direct coupling to the Higgs or
Stop fields. This will enable us to derive the electromagnetic fields
from EWPT bubble collisions as in Refs.\cite{kv95,ae98,cst00}. In
the present paper, however, we derive the electromagnetic fields
produced in the EWPT via bubble nucleation before collisions, which
has not been considered previously. For collisions a direction in space
is singled out, so that the form $W(r,t),A(r,t)$ cannot be used. This
is a topic for future work. Also, fermions contribute to the electric current,
and fermion fields will be included in future work.
The solution for $A^{em}$ produced during
nucleation with the assumptions of the present section are found in the 
following section.

\section{I-Spin Ansatz and Electromagnetic Field Creation During Nucleation.}

   In the present work we make use of the gauge fields gauge conditions to
reduce the partial differential equations, Eqs(\ref{eomW},\ref{eomA}), 
to ordinary differential equations. The philosophy is to derive the $W^{\pm}$
and $A^{em}$ fields as a function of time at a fixed r. Since from the
general structure of the equations we expect at time t that the bubble
wall will be at $r\; =\; r_w\;\simeq \;t$, we are mainly interested in the 
nature of
the fields near r = $r_w$. As we shall see, since the solutons are 
modified instanton-like in nature, the most significant region for magnetic
field creation for both nucleation, and for collisions, will be at the
bubble walls. 

   The solutions must be independent of the choice of gauge; however,
the choice of gauge is important for our work.  First we note that if we 
use the Lorentz gauge, as in our recent work on QCDPT bubble 
nucleation\cite{kwj04}, the method discussed below for deriving e.o.m. for 
$W(r,t)$ and $A(r,t)$  does not work. It is not that the solutions are not
correct, it is simply that the resulting e.o.m. is essentially the Lorentz 
gauge condition itself. There is no new physics.

  We use the Coulomb gauge, which is consistent with spherical spatial symmetry
and the forms $W(r,t)$ and $A(r,t)$:
\beq
\label{coulgauge}
   \sum_{j=1}^{3} \partial_j W^j &=& \sum_{j=1}^{3} \partial_j A^j = 0
 \;\; {\rm or} \nonumber \\
    r\partial_r W(r,t) + 3 W(r,t) &=& 0\;,  
\eeq
with solutions
\beq
\label{gsoln}
    W(r,t) &=& \frac{W_r(t)}{r^3}\;\; A(r,t) = \frac{A_r(t)}{r^3}\; .
\eeq

   From Eq.(\ref{eomW}) and the gauge condition (\ref{coulgauge}), 
one obtains differential equations for $W^t(r,t)$ and $W^j(r,t)$ 
(with the notation of $W^{t,j}$ for $W^\nu$, with $\nu=4$,$\nu=j=(1,2,3)$,
respectively). 
By combining them one finds that the Higgs and Stop fields are disconnected,
and obtains an e.o.m. for the functions $W_r(t)$ and $A_r(t)$
\beq
\label{eomWt}
  W_r''(t) - \frac{3t}{r^2} W_r'(t) +\frac{3}{r^2} W_r(t) +g\frac{t^2-r^2}
{r^3}W_r(t) \\
   (\frac{1}{t}W_r'(t)-\frac{3}{r^2}W_r(t)) &=& 0 \nonumber\\
\label{eomAt}
 A_r''(t) - \frac{3t}{r^2} A_r'(t) +\frac{3}{r^2} A_r(t) +G\frac{t^2-r^2}{r^3}
W_r(t) \\
 (\frac{1}{t}W_r'(t)-\frac{3}{r^2}W_r(t)) & =& 0 \nonumber
\eeq

We proceed by 1) finding initial conditions and numerical solutions to 
Eq.(\ref{eomWt}) for W(t) for a series of r-values, 2) fitting a function 
to these  values, and 3) finding the function
\beq
\label{H(t)}
      H_r(t) &=& \frac{t^2-r^2}{r^3}W_r(t)(\frac{1}{t}W_r'(t)-\frac{3}{r^2}
W_r(t)) \; ,
\eeq
which is used in Eq.(\ref{eomAt}) to obtain an approximate solution for
$A_r(t)$ and thereby A(r,t), using Eq.(\ref{gsoln}).

In Figure 1 W(t) is given for various values of r,
and the time for the creation of the bubble wall is clearly seen. In
Figure 2 similar results are shown for A(t).
Note that $A^{em}$ has an instanton-like behavior 
at the bubble wall surface in that it is infinite if $\rho \rightarrow 0$, 
and the effective
denominator is similar to that of an instanton near the wall, 
\beq
\label{20}
             A(r,t) &=& \frac{A_W}{((r^2 - t^2) +\zeta^2)^2} \; .
\eeq 
Away from the surface A(r,t) becomes
smaller than this instanton-like solution.   

\begin{figure}[ht]
\epsfig{file=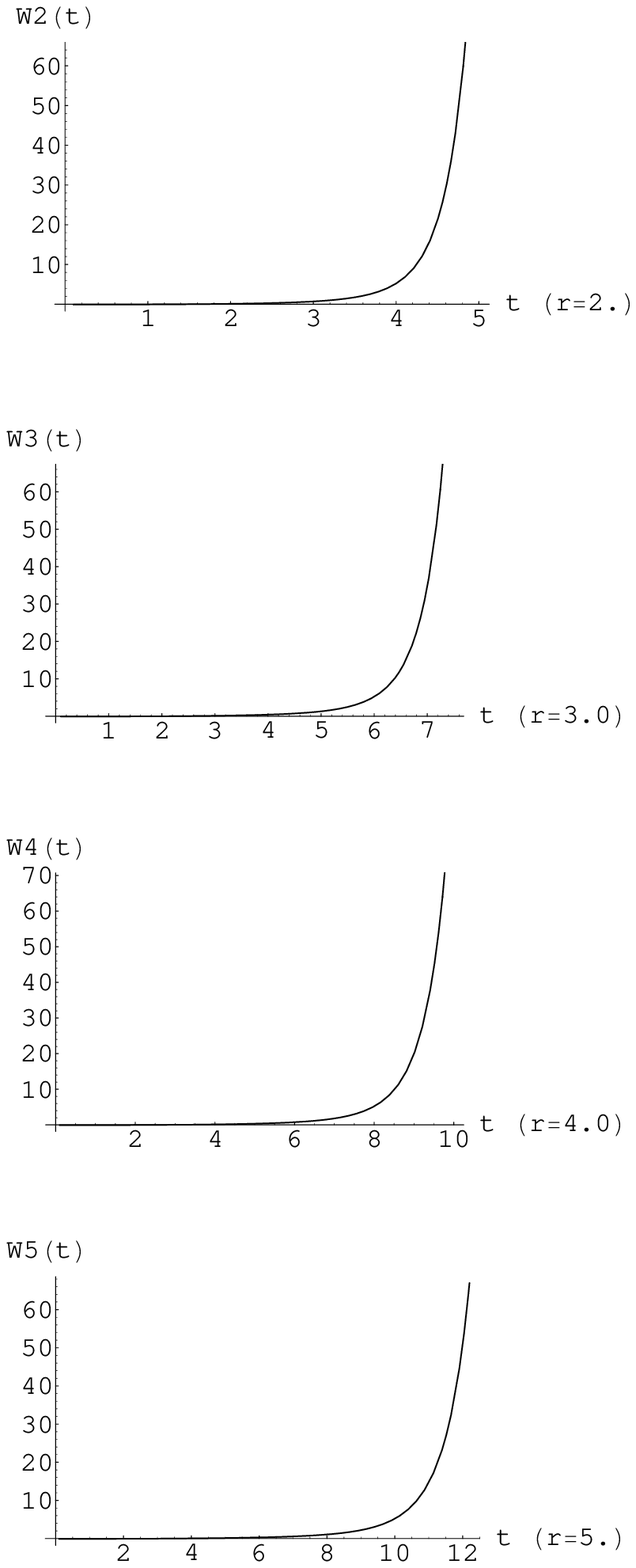,height=15cm,width=6.cm}
\caption{The function $W_r(t)$ for various values of r}
{\label{Fig.1}}
\end{figure}
\vspace{5cm}

\begin{figure}[ht]
\epsfig{file=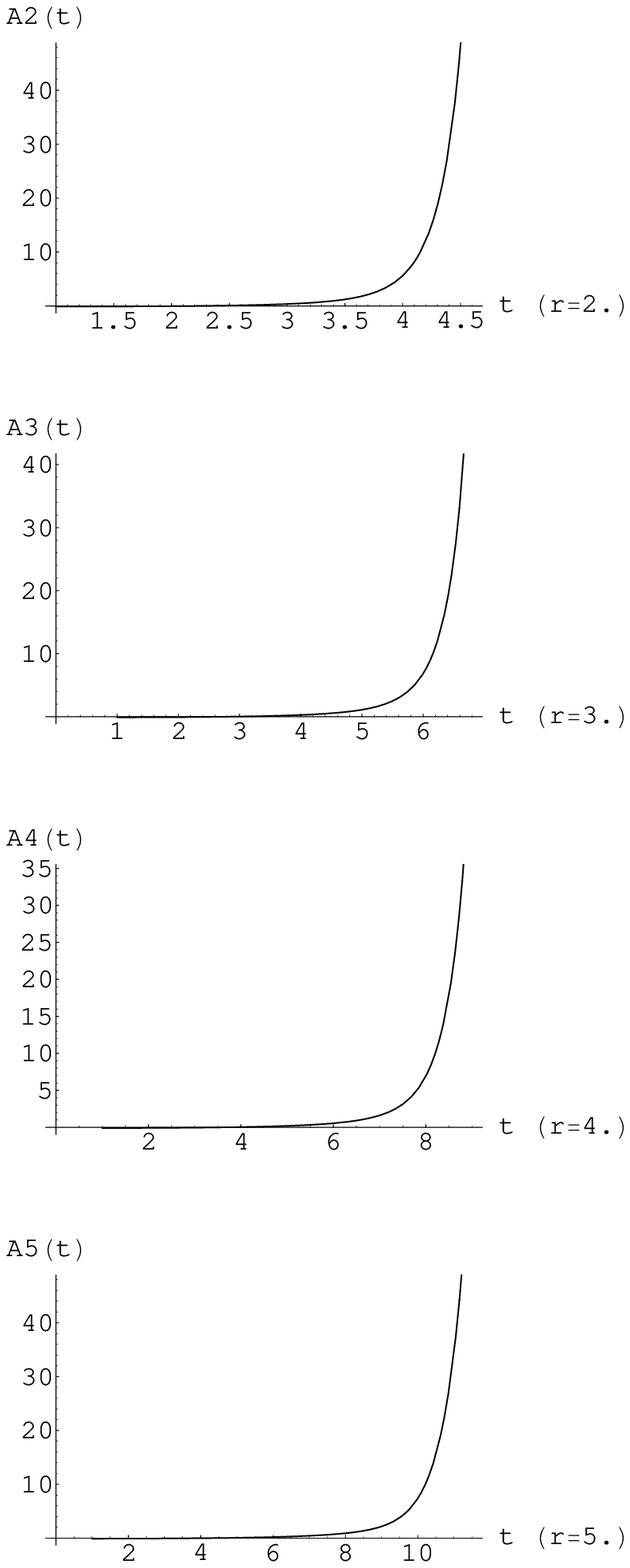,height=15cm,width=6.cm}
\caption{The function $A_r(t)$  for various values of r}
{\label{Fig.2}}
\end{figure}

   From Figure 2 one observes that the time at which one reaches the radius 
of the wall bubble is given approximately by
\beq
\label{26}
             t & \simeq & 2r \; ,
\eeq
from which we obtain the nucleation velocity of the bubble wall.
\beq
\label{27}
             v^{wall} & \simeq & \frac{c}{2} \; .
\eeq
This will be an important result in our future work on magnetic field 
generation
and evolution. In our present work $A^{em}_{\nu} \sim x_{\nu}A(r,t)$, 
so electric
fields but no magnetic fields are created. This work can provide the initial
conditions for EWPT collisions in which magnetic fields are created. 
Preliminary work on such bubble collisions has been carried out\cite{jkhhs04}.

\section{Conclusions}

   We have formulated the coupled equations of motion for the electroweak
MSSM with a Lagrangian that adds the right-handed Stop field terms to the 
Standard Model.
In this model a first order EWPT can occur with satisfactory baryogenesis.
By using an I-spin ansatz for spherically symmetric bubble nucleation we
were able to derive a Maxwell-like equation of motion for the electromagnetic
field with the current given by the electrically charged gauge fields, 
$W^{\pm}$. Moreover,
by treating the $W^t$ and $W^j$ components of $W^\nu$ separately we were able 
to decouple the equations for the $W^{\pm}$ fields from the other gauge 
fields,and also the Higgs and Stop fields, and obtain the current for the 
electromagnetic field.

In the present paper we derived solutions for the electromagnetic
field caused by EWPT bubble nucleation, using a Coulomb gauge condition
to obtain ordinary differential equations, from which we found instanton-like
solutions for the electromagnetic field in the region of the bubble wall.
Although this is a very limited physical problem, it explores new physics
which can arise from nucleation before collisions starting from a MSSM
electroweak Lagrangian. In our future work we will also examine EWPT bubble 
collisions, include fermion fields,  and examine possible predictions of 
galactic and extra-galactic magnetic structures.  We stress that during 
nucleation EWPT bubbles generate $A^{em}_{\nu}$, which must be included 
in the electromagnetic field arising from EWPT bubble collisions, and that
the length scale of the fields is that of the entire bubble, rather than
random fields within the bubble, which could be of importance in seeding
galactic and extra-galactic magnetic structures.

\vspace{3mm}

\Large{{\bf Acknowledgements}}\\
\normalsize
This work was supported in part by the NSF grant PHY-00070888, in part 
by the DOE contracts W-7405-ENG-36 and DE-FG02-97ER41014. 
The authors thank Prof. W.Y. Pauchy Hwang and Dr. S. Walawalkar for helpful 
discussions, and Los Alamos National Laboratory for hospitality.
\vspace{2cm}



\end{document}